\documentstyle[12pt]{article}

\makeatletter
\def\lesssim{\mathrel{\mathpalette\vereq<}}
\def\vereq#1#2{\lower3pt\vbox{\baselineskip1.5pt \lineskip1.5pt
\ialign{$\m@th#1\hfill##\hfil$\crcr#2\crcr\sim\crcr}}}

\def\gtrsim{\mathrel{\mathpalette\vereq>}}
\makeatother

\newcommand{\beq}{\begin{equation}}
\newcommand{\eeq}{\end{equation}}

\newcommand{\remove}[1]{}

\begin{document}
\begin{titlepage}
\begin{center}
\today     \hfill    LBL-36598 \\

\vskip .5in

{\large \bf Realistic Models with a Light U(1) Gauge Boson\\
Coupled to Baryon Number}\footnote{This
work was supported by the Director, Office of Energy Research,
Office of High Energy and Nuclear Physics, Division of High
Energy Physics of the U.S. Department of Energy
under Contract DE-AC03-76SF00098.}

\vskip 0.5in

Christopher D. Carone  and Hitoshi Murayama\footnote{On leave of absence from
Department of Physics, Tohoku University, Sendai, 980 Japan.}

{\em Theoretical Physics Group\\
     Lawrence Berkeley Laboratory\\
     University of California\\
     Berkeley, California 94720}

\end{center}

\vskip .3in

\begin{abstract}
We recently showed that a new gauge boson $\gamma_B$ coupling only
to baryon number is phenomenologically allowed, even if $m_B<m_Z$ and
$\alpha_B \approx 0.2$. In our previous work we assumed that kinetic mixing
between the baryon number and hypercharge gauge bosons (via an $F^{\mu\nu}_B
F_{\mu\nu}^Y$ term) was small enough to evade constraints from precision
electroweak measurements.  In this paper we propose a class of models in
which this term is naturally absent above the electroweak
scale. We show that the generation of a mixing term through radiative
corrections in the low-energy effective theory does not lead to conflict
with precision electroweak measurements and may provide a leptonic
signal for models of this type at an upgraded Tevatron.
\end{abstract}

\end{titlepage}
\renewcommand{\thepage}{\roman{page}}
\setcounter{page}{2}
\mbox{ }

\vskip 1in

\begin{center}
{\bf Disclaimer}
\end{center}

\vskip .2in

\begin{scriptsize}
\begin{quotation}
This document was prepared as an account of work sponsored by the United
States Government. While this document is believed to contain correct
information, neither the United States Government nor any agency
thereof, nor The Regents of the University of California, nor any of their
employees, makes any warranty, express or implied, or assumes any legal
liability or responsibility for the accuracy, completeness, or usefulness
of any information, apparatus, product, or process disclosed, or represents
that its use would not infringe privately owned rights.  Reference herein
to any specific commercial products process, or service by its trade name,
trademark, manufacturer, or otherwise, does not necessarily constitute or
imply its endorsement, recommendation, or favoring by the United States
Government or any agency thereof, or The Regents of the University of
California.  The views and opinions of authors expressed herein do not
necessarily state or reflect those of the United States Government or any
agency thereof or The Regents of the University of California and shall
not be used for advertising or product endorsement purposes.
\end{quotation}
\end{scriptsize}

\vskip 2in

\begin{center}
\begin{small}
{\it Lawrence Berkeley Laboratory is an equal opportunity employer.}
\end{small}
\end{center}

\newpage
\renewcommand{\thepage}{\arabic{page}}
\setcounter{page}{1}
\section{Introduction}\label{sec:intro}

In a recent paper, we considered the phenomenology of a light U(1)
gauge boson $\gamma_B$ that couples only to baryon number
\cite{carmur1}.  We assumed that the new U(1) gauge symmetry is
spontaneously broken, and that the $\gamma_B$ mass $m_B$ is smaller than
$m_Z$. Nevertheless, we showed that this new gauge boson could remain
undetected, even if the coupling $\alpha_B$ were comparable to $\alpha_{{\rm
strong}}$ \cite{carmur1,sddb}. Since the $\gamma_B$ boson couples only to
quarks, any process that is relevant to $\gamma_B$ detection also has a
significant contribution from QCD.  Thus, a typical $\gamma_B$ boson with
$m_B=50$ GeV and $\alpha_B = 0.1$, can remain undetected by `hiding' in
the large QCD background.  Since the $\gamma_B$ boson couples only
to quarks, it is difficult to detect, just like the more familiar
example of a light gluino in supersymmetric models \cite{susy}.

One of the assumptions in our original analysis \cite{carmur1} was
that mixing between the $\gamma_B$ and electroweak gauge bosons was
negligible. Mass mixing is not present because we assume that there
are no Higgs bosons that carry both baryon number and
electroweak quantum numbers.  However, there is a possible off-diagonal
kinetic term that mixes the U(1)$_B$ and U(1)$_Y$ gauge fields:
\beq
{\cal L}_{kin}=-\frac{1}{4} \left( F^{\mu\nu}_Y F_{\mu\nu}^Y
+ 2 c\, F^{\mu\nu}_B F_{\mu\nu}^Y + F^{\mu\nu}_B F_{\mu\nu}^B
\right) .
\label{eq:mixdef}
\eeq
Here the $F^{\mu\nu}$ are gauge field strength tensors,
and $c$ is an undetermined coupling constant.  Clearly, $c$ must be
quite small so that the kinetic mixing does not conflict with
precision electroweak measurements.  Although the phenomenology
of the $\gamma_B$ is specified within the three dimensional parameter
space $m_B$-$\alpha_B$-$c$, any realistic model must reside within
the narrow region $|c|<c_0$, where $c_0\ll 1$ can be determined
from the precision electroweak constraints.  Thus, in our previous work
\cite{carmur1}, we described the $\gamma_B$ phenomenology in terms
of an effectively two-dimensional parameter space,
the $m_B$-$\alpha_B$ plane at $c\approx 0$.

The natural question that remains to be answered is whether there
are any models in which $c$ is naturally small enough to satisfy the
experimental constraints.  Our previous results would be greatly
undermined if they were relevant only to models in which the coupling $c$
required fine-tuning at the electroweak scale.  In this paper, we will
describe a class of models in which this kinetic mixing
term is absent above some scale $\Lambda$ that we assume is not much
greater than the top quark mass. Below $\Lambda$, a kinetic mixing term
is generated only through radiative corrections, so that $c(\Lambda)=0$
but $c(\mu)\neq 0$ for $\mu < \Lambda$.  We will show that if
$200 \mbox{ GeV} \lesssim \Lambda \lesssim 1.3$ TeV, $c(\mu)$ never
becomes large enough in the low-energy theory to conflict with precision
electroweak measurements, even when $\alpha_B$ is as large as $0.1$.
We then present a model that satisfies this boundary condition. In
addition, we show that a mixing term small enough to satisfy the current
experimental constraints can nonetheless provide us with a possible
signal for the $\gamma_B$ in the Drell-Yan dilepton differential cross
sections at hadron colliders.  This signal could be within the reach of
the Fermilab Tevatron with the main injector and a luminosity upgrade.

The paper is organized as follows.  In the next section we discuss the
phenomenological constraints on the kinetic mixing term from precision
electroweak measurements.  We show that these constraints can be
satisfied if the scale $\Lambda$ at which the mixing vanishes is just
above the electroweak scale. In section 3, we present a model with
gauged baryon number in which the kinetic mixing term is naturally
absent above $\Lambda$.  In section 4 we discuss a leptonic signature of
the $\gamma_B$ in Drell-Yan dilepton production at hadron colliders.  In
the final section we summarize our conclusions.

\section{Mixing Constraints}
\setcounter{equation}{0}

To study the effects of the kinetic mixing term, we could
redefine the gauge fields so that the kinetic terms in
the new basis are diagonal and conventionally normalized \cite{ntet}.
However, since we know {\it a priori} that the coupling $c$
is much less than 1, it is more convenient to treat the mixing
term in Eq.~(\ref{eq:mixdef}) as a new perturbative interaction.
The Feynman rule corresponding to the $\gamma_B$-photon vertex
is
\beq
-i c_\gamma \cos\theta_w (p^2 g^{\mu\nu}-p^\mu p^\nu)
\eeq
and for the $\gamma_B$-$Z^0$ vertex is
\beq
i c_Z \sin\theta_w (p^2 g^{\mu\nu}-p^\mu p^\nu) .
\eeq
Note that $c_\gamma=c_Z=c$ above the electroweak scale, but
$c_\gamma$ and $c_Z$ run differently in the low-energy
effective theory below $m_{{\rm top}}$.  This can be seen in Fig.~1.
If we assume that both $c_Z$ and $c_\gamma$ vanish at the
scale $\Lambda \approx m_{{\rm top}}$, we can run the couplings down to
lower energies to determine the magnitude of mixing that is characteristic
of purely radiative effects.  This will give us a useful point of reference
when we consider the relevant experimental constraints.  If $\Lambda$ were
less than $m_{{\rm top}}$, the new physics that is associated with this
scale would have already been seen in accelerator experiments.  The extent
to which $\Lambda$ can be significantly larger than $m_{{\rm top}}$ will
be considered later in this section.

The couplings $c_Z$ and $c_\gamma$ are renormalized by the one quark-loop
diagrams that connect the $\gamma_B$ to either the photon or $Z$.  From
these diagrams, we obtain the following one-loop
renormalization group equations
\beq
\mu\frac{\partial}{\partial \mu} c_\gamma(\mu) = - \frac{2}{9\pi}
\frac{\sqrt{\alpha_B\alpha}}{c_w}\,\left[2N_u-N_d\right] ,
\label{eq:runcg}
\eeq
and
\beq
\mu\frac{\partial}{\partial \mu} c_Z(\mu) = - \frac{1}{18\pi}
\frac{\sqrt{\alpha_B\alpha}}{s^2_wc_w}\left[3(N_d-N_u)+4
(2N_u-N_d) s_w^2 \right] ,
\label{eq:runcz}
\eeq
where $s_w\,(c_w) = \sin\theta_w \, (\cos\theta_w)$, and $\alpha$
is the fine structure constant.  $N_u$ and $N_d$ are the number of
light quark flavors with charge $2/3$ and $-1/3$, respectively.
For $\Lambda=m_{{\rm top}}$, and $\alpha_B=0.1$, we find
\begin{center}
\begin{tabular}{cc}
$c_\gamma(m_Z) \approx 1.5 \times 10^{-3}$, &
$c_Z(m_Z) \approx 6.2 \times 10^{-3}$, \\
$c_\gamma(m_b) \approx 8.4 \times 10^{-3}$, &
$c_Z(m_b) \approx 3.5 \times 10^{-2}$, \\
\end{tabular}
\end{center}
where $m_b$ is the bottom quark mass.  As we cross the various quark
mass thresholds below $m_b$, the rate at which $c_\gamma$ and $c_Z$ run
becomes progressively smaller, until the running effectively stops below
the hadronic scale $\approx 1$ GeV.  It is our first job to determine
whether the estimates above are consistent with experimental limits
on $c_\gamma$ and $c_Z$.  Afterwards, we will return to the issue of
the running and determine an upper bound on the scale $\Lambda$.

\subsection{$\gamma_B$-$Z$ Mixing}

The most significant constraints on $c_Z(m_Z)$ are shown in Fig.~2.
We have considered the effects of the $\gamma_B$-$Z$ mixing on
the following experimental observables: the $Z$ mass, hadronic width,
and forward-backward asymmetries, and the neutral current $\nu N$ and
$e N$ deep inelastic scattering cross sections. We now consider each
of these in turn.

{\em $Z$ mass.} We determine the shift in the $Z$ mass by computing
the shift in the real part of the pole in neutral gauge boson
propagator. Thus, we set
\beq
\det \Gamma^{(2)}(p^2) = 0 ,
\eeq
where
\beq
\Gamma^{(2)}(p^2) = \left(\begin{array}{cc}
p^2-m_Z^2+i m_Z \Gamma_Z & - c_Z s_w p^2 \\
- c_Z s_w p^2 & p^2-m_B^2+i m_B \Gamma_B \end{array} \right) .
\eeq
We find
\beq
\frac{\Delta m_Z}{m_Z} \approx 0.116 \, c_Z^2
\frac{m_Z^2}{m_Z^2-m_B^2} .
\label{eq:dmz}
\eeq
Note that the effect of $\gamma_B$-$\gamma$ mixing appears at
${\cal O}(c_Z^2 c_\gamma^2)$, and is negligible.
This expression is valid provided that $m_B$ is not
too close to $m_Z$. We have checked that this approximation
is accurate if $|m_B-m_Z| \gtrsim 10$ GeV, which
holds over range of $m_B$ of interest to us in this paper.  Since $m_Z$
is taken as an input to determine other electroweak parameters, we require
that the shift in $m_Z$ does not spoil the consistency between the value
of $\sin^2 \hat{\theta}_w$ determined from the $Z$ decay asymmetries
(which we call $s^2$ below), and the value extracted from deep inelastic
scattering data.  The shift in $m_Z$ corresponding to the uncertainty
$\Delta s^2$ is given by
\beq
\frac{\Delta m_Z}{m_Z} = -\frac{1-2 s^2}{2 s^2 c^2} \Delta s^2 ,
\eeq
where $s^2 = 0.2317 \pm 0.0008$ \cite{pdg}.  Thus, we find
$\Delta m_Z / m_Z < 2.4 \times 10^{-3}$, requiring that the
shift in $\sin^2\hat{\theta}$ is no more than a two standard
deviation effect. The contour corresponding to this bound is plotted
in Fig.~2.

{\em Z hadronic width.}  In Ref.~\cite{carmur1} we computed
the contribution to the $Z$ hadronic width from (i) direct $\gamma_B$
production $Z\rightarrow q\overline{q}\gamma_B$, and (ii) the
$Zq\overline{q}$ vertex correction.  There is an additional
contribution to the hadronic width from the $Z$-$\gamma_B$
mixing that is given by
\beq
\frac{\Delta \Gamma_{had}}{\Gamma_{had}} \approx -1.194 \, c_Z
\sqrt{\alpha_B} \frac{m_Z^2}{m_Z^2-m_B^2} ,
\label{eq:newhad}
\eeq
Given that the uncertainty in the $Z$ hadronic width is 0.6\% at
two standard deviations \cite{pdg}, we obtain the contour shown in
Fig.~2.  Notice that the constraint that we obtain is weaker
for positive $c_Z$ due to cancelation between the contributions
discussed in Ref.~\cite{carmur1}, and the new contribution
given in Eq.~(\ref{eq:newhad}).  In either case, the hadronic width
places the tightest constraint on $c_Z(m_Z)$, roughly
$|c_Z(m_Z)| \lesssim 0.02$.

{\em Forward-Backward Asymmetries.} The $\gamma_B$-$Z$ mixing term
has the effect of slightly shifting the vector coupling of the $Z$
to quarks. Thus, there is a new contribution to the forward-backward
asymmetry $A_{FB}^{(0,q)}$ in $Z$ decay to $q\overline{q}$.  Since
the experimental uncertainty is smallest for $q=b$, we use
the two standard deviation uncertainty in $A_{FB}^{(0,b)}$ to
constrain our model.  We find that the new contribution is given by
\beq
\Delta A_{FB}^{(0,b)} \approx - 0.159 \, c_Z \sqrt{\alpha_B}
\frac{m_Z^2}{m_Z^2-m_B^2} ,
\eeq
while the measured value is 0.107$\pm$0.013 \cite{pdg}.  The
resulting bound is shown in Fig.~2.  Notice that this provides
a weaker constraint than those we obtained from consideration of
the $Z$ mass and width.

{\em Deep Inelastic Scattering.}  The constraints on $c_Z$ from
deep inelastic $\nu N$ scattering, and from parity-violating
$e N$ scattering are much weaker than the other constraints
that we have discussed, and are not shown in Fig.~2.  Deep
inelastic $\nu N$ scattering can be described in terms of the
parameters $\epsilon_{L(R)}$, defined by the effective four-fermion
interaction \cite{pdg}
\begin{equation}
-{\cal L}^{\nu N} =
\frac{G_F}{\sqrt{2}} \overline{\nu}
\gamma^\mu(1-\gamma^5) \nu  \sum_i \left[ \epsilon_L(i)
 \overline{q_i} \gamma_\mu(1-\gamma^5) q_i + \epsilon_R(i)
\overline{q_i} \gamma_\mu(1+\gamma^5) q_i \right]  ,
\end{equation}
where the sum is over quark flavors.  We find that the contribution
to the $\epsilon$ parameters from the $\gamma_B$-$Z$ mixing is given
by
\beq
\Delta \epsilon_{L(R)} \approx 0.766 \, c_Z \sqrt{\alpha_B}
\frac{q^2}{q^2-m_B^2} ,
\label{eq:epcon}
\eeq
where $-q^2 \approx 20$ GeV$^2$ is a typical squared momentum
transfer. The most accurately measured $\epsilon$ parameter is
$\epsilon_L(d) = -0.438 \pm 0.012$ \cite{pdg}.  To demonstrate that the
uncertainty in $\epsilon_L(d)$ provides only a weak constraint, we
evaluate Eq.~(\ref{eq:epcon}) for $m_B \approx 50$ GeV and
$\alpha_B \approx 0.1$.  We obtain the bound $|c_Z(q^2)| < 12.5$,
at two standard deviations, which gives us $|c_Z(m_Z)|< 12.5$, because
the contribution from the running is small. This is a much weaker
constraint than the others that we have considered.

Parity-violating $eN$ scattering can be described in terms of two
other parameters, $C_1$ and $C_2$, defined by the effective four-fermion
interaction \cite{pdg}
\beq
-{\cal L}^{e N} = - \frac{G_F}{\sqrt{2}} \sum_i \left[ C_{1i}
\overline{e} \gamma_\mu \gamma^5 e \overline{q_i} \gamma^\mu q_i
+ C_{2i} \overline{e} \gamma_\mu e
\overline{q_i} \gamma_\mu\gamma^5 q_i \right]  .
\eeq
The $\gamma_B$-$Z$ mixing contributes only to the parameter $C_{1i}$:
\beq
\Delta C_{1i} = - 1.533 \, c_Z \sqrt{\alpha_B}
\frac{q^2}{q^2-m_B^2} .
\eeq
The parameter measured with the least experimental uncertainty is
$C_{1d}=0.359 \pm 0.041$ \cite{pdg}.  If we again assume that $m_B=50$
GeV and $\alpha_B=0.1$, then the bound on $c_Z(m_Z)$ following from the
two standard deviation uncertainty in $C_{1d}$ is $|c_Z(m_Z)|< 21.2$.
This is even weaker than the constraint we obtained from $\nu N$
scattering.  Note that there are no further constraints on
$\Delta C_{1i}$ from the measurements of atomic parity
violation because this process involves zero momentum transfer,
where the kinetic mixing vanishes.

What we have seen is that the $Z$-pole observables place the
tightest constraints on the mixing parameter $c_Z$, while
deep inelastic scattering measurements do not provide any
further constraints.  Thus, if $|c_Z(M_Z)|<0.02$,
we are not likely to encounter any problems with the precision
electroweak measurements that we have considered in this section.

\subsection{$\gamma_B$-$\gamma$ Mixing}

The coupling $c_\gamma$ has its most significant effect on
a different set of observables.  Below we consider the effect of
the $\gamma_B$-$\gamma$ mixing on the cross section for
$e^+e^-\rightarrow hadrons$, and on the anomalous
magnetic moments of the electron and muon.

{$e^+e^-\rightarrow hadrons$.} The most important constraint on
$c_\gamma$ comes from the additional contribution to $R$, the ratio
$\sigma(e^+e^-\rightarrow \mbox{hadrons})/
\sigma(e^+e^-\rightarrow \mu^+\mu^-)$.  We find
\begin{equation}
\frac{\Delta R}{R} =
-1.803 \, c_\gamma \sqrt{\alpha_B}
\frac{s(s-m_B^2)}{(s-m_B^2)^2+m_B^2 \Gamma_B^2}
+8.938 \,c_\gamma^2 \, \frac{s^2}{(s-m_B^2)^2+m_B^2 \Gamma_B^2} ,
\end{equation}
where $\sqrt{s}$ is the center of mass energy, and
$\Gamma_B = \frac{1}{9} N_F \alpha_B m_B$ is the $\gamma_B$ width,
with $N_F = 5$ for the range of $m_B$ of interest.
Notice that the nonstandard contribution to $R$ is maximized only in the
vicinity of $s\approx m_B^2$. For any $m_B$ of interest, we can
constrain $c_\gamma$ by considering the two standard deviation uncertainty
in the value of $R$ measured at $\sqrt{s}\approx m_B$.  The results are shown
in Fig.~3, based on the cumulative data on $R$ taken at various values of
$\sqrt{s}$ and compiled by the Particle Data Group \cite{pdg}.
Since there are values of $\sqrt{s}$ that have not been studied, the
constraints on $c_\gamma$ are strongest when $m_B$ happens to coincide with
a value of $\sqrt{s}$ at which there is an experimental data point available.
Roughly speaking, the allowed region of Fig.~3 corresponds to
$|c_\gamma(m_B)|<0.01$. However, it is clear that the constraint can be
significantly weaker if $m_B$ happens to lie at point where less
data are available.

{\em Anomalous magnetic moments.}  At the very lowest energies, we can
constrain $c_\gamma$ by the effect of the mixing on the anomalous
magnetic moments of the electron and muon.  Since this provides a much
weaker constraint than the one we obtained from $R$, we do not show the
result in Fig.~3.  We find that the nonstandard
contribution to the anomalous magnetic moment $a=(g-2)/2$ is given by
\beq
\Delta a = c_\gamma^2 \, c_w^2 \, \frac{\alpha}{\pi} \, I(r) ,
\eeq
where
\beq
I(r)=\frac{1}{2}-r+\frac{r}{2} (r-2) \ln(r)-\frac{r}{2}
\frac{(r^2-4r+2)}{\sqrt{r^2-4r}}\ln\frac{r+\sqrt{r^2-4r}}
{r-\sqrt{r^2-4r}} ,
\eeq
and $r=m_B^2/m^2_{lepton}$. Since $r$ is large, we use the
asymptotic form $I(r)\approx 1/(3r)$. Then the limit on $c_\gamma$
corresponding to a two standard deviation uncertainty in
the anomalous magnetic moment is
\beq
c_\gamma(m_\mu) < 0.050 \, \left(\frac{m_B}{\mbox{GeV}} \right)
\eeq
for the muon, and
\beq
c_\gamma(m_e) < 0.360 \, \left(\frac{m_B}{\mbox{GeV}} \right)
\eeq
for the electron.  Thus, the constraint on $c_\gamma$ for
$m_B > 20$ GeV is roughly two orders of magnitude weaker than
the constraints that we obtained from $R$.

\subsection{The Scale $\Lambda$}

It should now be clear that our original estimate of the
sizes of $c_\gamma(\mu)$ and $c_Z(\mu)$ fall within the
bounds that we have obtained from consideration of
precision electroweak measurements.  Recall that the estimate that
we presented at the beginning of this section was for
$c_\gamma(\Lambda)=c_Z(\Lambda)=0$ at $\Lambda=m_{{\rm top}}$.
We will now determine how high we can push up $\Lambda$ before we have
unambiguous conflict with the experimental constraints.  Using the
approximate bound that we obtained for $c_Z$, $c_Z(m_Z)<0.02$,
and assuming $m_{{\rm top}}\approx 175$ GeV, we find
\beq
\Lambda < 1.3 \,\,\mbox{TeV}
\label{eq:lbnd}
\eeq
from Eqs.~(\ref{eq:runcg}) and (\ref{eq:runcz}) with $\alpha_B=0.1$.  We
can place comparable bounds on $\Lambda$ from the constraints
on $c_\gamma$, but the precise result depends crucially on the choice
for $m_B$, as one can see from Fig.~3.  What is interesting about
Eq.~(\ref{eq:lbnd}) is that it implies that the scale of new physics lies
at relatively low energies, just above the electroweak scale. We will now
show that there is a class of models that satisfy the desired boundary
condition at $\mu=\Lambda$.

\section{Models with Naturally Small Kinetic Mixing}
\setcounter{equation}{0}

In this section, we present a simple model with gauged baryon number
that naturally satisfies the boundary condition
$c_\gamma(\Lambda)=c_Z(\Lambda)=0$
with $\Lambda < 1.3$~TeV.  There are two ingredients that are of central
importance in the class of models that have small kinetic mixing below
the electroweak scale. (1) In the full theory, at high energies, the
kinetic mixing term is forbidden by gauge invariance.  This is the case,
for example, if we embed one of the U(1)s in a larger nonabelian
group. The mixing term remains vanishing down to the scale at which the
gauge symmetry breaks to $G\times$U(1)$_B\times$U(1)$_Y$, where $G$
contains the remaining gauge structure of the theory.  (2) Beneath this
symmetry breaking scale, the one-loop diagram that connects the
$\gamma_B$ to the hypercharge gauge boson vanishes identically, so that
$c_\gamma$ and $c_Z$ do not run.  This places a constraint on the
particle content beneath the symmetry breaking scale,
\beq
\mbox{Tr}\,(B\,Y) = 0
\label{eq:orth}
\eeq
where $B$ and $Y$ are baryon number and hypercharge matrices,
in the basis spanning the entire particle content of
the theory.   When we go to lower energies and the heaviest
particle that contributes to Eq.~(\ref{eq:orth}) is integrated out, we will
generate mixing through radiative corrections, in the way
described quantitatively in section~2.

In what follows, we will present one example of a model with gauged
baryon number that is `realistic' in the following sense: (i) the kinetic
mixing is naturally small below the electroweak scale, (ii) there is a
natural mechanism for generating the cosmic baryon asymmetry, and (iii)
proton decay is forbidden (up to the usual non-perturbative effects) even
though U(1)$_B$ is spontaneously broken.  It is not our goal to study
every aspect of the phenomenology of this particular model, but rather
to demonstrate by example that it is possible to construct models with
the features (i), (ii),  and (iii). In addition, we show that there are
new fermions in the models of interest that appear in chiral representations
of SU(2)$_L \times$ U(1)$_Y$;  these fermions develop electroweak scale
masses $\gtrsim m_{{\rm top}}$, which ensures that the scale $\Lambda$
is not too far above the electroweak scale.  Moreover, this  implies
that detection of the new fermions in this class of models is likely at
an upgraded Tevatron or at the LHC.

\subsection{A Model}
The gauge structure of the model is
\begin{displaymath}
\mbox{SU(3)}_{C} \times \mbox{SU(2)}_L \times \mbox{U(1)}_Y
	\times \mbox{SU(4)}_{H} ,
\end{displaymath}
where SU(4)$_H$ is a horizontal symmetry.  In addition to the
ordinary three families of the standard model, $f^i$~$(i=1,2,3)$,
we assume there is a fourth family $F$; the horizontal symmetry
acts only on the quarks in the four families, which together
transform as a {\bf 4} under the SU(4)$_H$.  The U(1)$_B$ gauge group
is embedded into SU(4)$_H$ as
\begin{equation}
	B = \left( \begin{array}{cccc}
			1/3 &   &   &  \\
			  & 1/3 &   &  \\
			  &   & 1/3 &  \\
			  &   &   & -1
		   \end{array} \right) .
\label{Bdef}
\end{equation}
While SU(4)$_H$ is broken at some high scale $M_H$, as we discuss below,
the U(1)$_B$ subgroup remains unbroken down to the electroweak scale.
It is easy to verify that the particle content and quantum number
assignments render the model anomaly-free.

We will also assume that there are right-handed neutrinos in each of
the families.  The right-handed neutrinos in the ordinary families
acquire Majorana masses at a high scale $M_N$, while the one in the
fourth family does not\footnote{This is natural if there is another
global or local SU(4)$_H$ acting on the leptons, and if lepton number
embedded as $L = \mbox{diag} (1, 1, 1, -3)$ in SU(4)$_H$ is broken by an
order parameter with $L=-2$ (or {\bf 10} under SU(4)$_H$). However,
none of the conclusions in this paper depends on whether there exists a
horizontal symmetry for the leptons or not.}. The fourth family neutrino
develops an electroweak scale Dirac mass, so that the constraint from
the invisible decay width of the $Z$ is evaded.  The Majorana masses
for the right-handed neutrinos in the first three families are
crucial to the  baryogenesis scenario that we present in the next
subsection. An interesting choice for the Majorana mass scale
is $M_N \sim 10^{10}$--$10^{12}$~GeV, which is consistent with the MSW
solution to the solar neutrino problem, and the possibility that
$\nu_\tau$ is a hot dark matter particle \cite{review}.

The horizontal symmetry SU(4)$_H$ is broken at a
scale $M_H$ down to U(1)$_B$.  One can imagine that the symmetry breaks
in one step if a number of adjoint Higgs bosons, that we will generically
call $\Phi$, develop vacuum expectation values (VEVs) at the scale $M_H$.
However, it is possible to generate a hierarchy of fermion
Yukawa couplings if we break the symmetry sequentially, in the
presence of additional vector-like fermions \cite{Nielsen}.  The basic
idea is as follows:  We first introduce vector-like fermions $\Psi$ with mass
$M$ that transform as {\bf 4}s under SU(4)$_H$, and assume that
the electroweak Higgs boson $H$ is a singlet under the horizontal
symmetry. Suppose that the $\Psi$ have Yukawa interactions like ${\cal L} =
\bar{q}_L \Phi \Psi + \overline{\Psi} H u_R + M \overline{\Psi} \Psi$.
Then, when the vector-like fermions are integrated out below the scale
$M$, we obtain dimension 5 operators of the form
\beq
\frac{1}{M}\overline{q}_L H \Phi u_R\ .
\label{eq:massop}
\eeq
Notice when the $\Phi$s develop VEVs, operators like the one in
Eq.~(\ref{eq:massop}) generate Yukawa couplings in the low-energy theory.
Now imagine that the horizontal symmetry is broken first to
SU(3)$_H \times$ U(1)$_B$ at the scale $M_H$, and then the SU(3)$_H$ is
broken sequentially at lower scales down to nothing.  Then the dimension
5 operators that we have described can give rise to a hierarchical pattern
of Yukawa couplings.  A detailed analysis of the fermion mass matrix in
models with horizontal symmetry breaking is beyond the scope of this paper,
and we refer the interested reader to the literature \cite{horizontal}.

The low-energy particle content of our model below both $M_H$ and $M_N$ is
listed in the Table~1. Here, $B$ refers to the gauge quantum number
under U(1)$_B$, while $L$ is an effective non-anomalous global symmetry
below $M_{N}$.  $(B-L)_{\rm extra}$ is another non-anomalous global
symmetry acting on the particles in the fourth family.

\begin{table}
\caption[content]{Particle content below the horizontal symmetry breaking
scale $M_{H}$ and the right-handed neutrino masses $M_N$.}
\centerline{
\begin{tabular}{|c|c|cccccc|}
\hline
      &particle & SU(3)$_C$ & SU(2)$_L$ & U(1)$_Y$ & $B$ & $L$ &
	$(B-L)_{\rm extra}$\\
\hline
    &$q^i_L$ & 3 & 2 & 1/6 & 1/3 & 0 & 0\\
ordinary families&$u^i_R$ & 3 & 1 & 2/3 & 1/3 & 0 & 0\\
$f_L^i$, $f_R^i$&$d^i_R$ & 3 & 1 & $-$1/3& 1/3 & 0 & 0\\
$(i=1,2,3)$&$l^i_L$ & 1 & 2 & $-$1/2&  0  & 1 & 0\\
    &$e^i_R$ & 1 & 1 & $-$1  &  0  & 1 & 0\\
\hline
    &$Q_L$ & 3 & 2 & 1/6 &$-$1 & 0 & $-$1\\
    &$U_R$ & 3 & 1 & 2/3 &$-$1 & 0 & $-$1\\
extra family&$D_R$ & 3 & 1 & $-$1/3&$-$1 & 0 & $-$1\\
$F_L$, $F_R$&$L_L$ & 1 & 2 & $-$1/2& 0 &$-$3 & +3\\
    &$E_R$ & 1 & 1 & $-$1  & 0 &$-$3 & +3\\
    &$N_R$ & 1 & 1 &  0  & 0 &$-$3 & +3\\
\hline
\end{tabular}
}
\end{table}

It is easy to see that the kinetic mixing remains vanishing down to the
weak scale. Above $M_{H}$, the mixing is not allowed because U(1)$_B$ is
embedded into the non-abelian group SU(4)$_{H}$.  This implies that the
orthogonality condition (\ref{eq:orth}) is satisfied by the particle
content of the full theory.  As we cross $M_{H}$, presumably all fields
whose mass terms are allowed by the gauge symmetry decouple, but
the particles listed
in the table do not because they belong to chiral representations of
the gauge group below $M_H$. One can easily check
that the orthogonality between $Y$ and $B$ remains true below $M_{H}$ as
well given the particle content in Table~1.
The mixing term is only generated below the masses of the
particles in the extra family (which we will refer to generically as
$m_F$) which originate from electroweak symmetry breaking. Therefore,
the mixing term remains vanishing down to the weak scale, {\it i.e.},\/
$\Lambda = m_F \sim m_{\rm top}$, and the
boundary condition discussed in the previous section is naturally
achieved.

\subsection{Baryogenesis and Proton Stability}

It is natural to wonder how a cosmic baryon asymmetry can be generated
in a model in which baryon number is a gauge symmetry at high energies.
On the other hand, it is natural to worry about proton decay considering
that the baryon number gauge symmetry is spontaneously broken at low
energies.  We address these two issues in this subsection.  With regard to
baryogenesis, we will show that the generation of a lepton number
asymmetry from the decay of the right-handed Majorana neutrinos in our
model can lead to a nonvanishing baryon number for particles from the
ordinary three families, even though the total baryon number of the
universe remains zero. We describe this mechanism in some detail, as well
as other relevant cosmological issues.  Afterwards, we demonstrate that
proton decay is forbidden in the model, apart from the electroweak
non-perturbative effects.

The first step in baryogenesis is that a lepton asymmetry is generated
from the CP-violating decays of the right-handed Majorana neutrinos.
If we take into account the effect of the electroweak anomaly at a
temperature at or above the electroweak phase transition \cite{KRS},
chemical equilibrium leads to non-vanishing lepton and baryon numbers
in both the ordinary and extra families. Finally, the quarks in the extra
family decay into those of the ordinary families, so that a cosmic
over-density of fourth-generation particles is avoided.  The first step
is exactly the one proposed in Ref.~\cite{FY} (see also \cite{nhr}
for the supersymmetric case.). The Yukawa interactions
coupling the right-handed neutrinos to the lepton doublets
violate CP in general; thus, the decay of the right-handed neutrinos
can generate a net lepton asymmetry.

The analysis of chemical equilibrium including the electroweak anomaly
effect is more complicated than in the Minimal Standard Model \cite{HT}.
Note that the total lepton number
\begin{equation}
L = (N_l + N_e) - 3 (N_L + N_E + N_N)
\end{equation}
is non-anomalous contrary to the minimal case, and is
broken only by the small Majorana masses of the left-handed neutrinos
generated by the seesaw mechanism \cite{seesaw}. In addition, there are
the non-anomalous conserved quantum numbers
\begin{equation}
B = \frac{1}{3} (N_q + N_u + N_d) - (N_Q + N_U + N_D),
\end{equation}
and
\begin{equation}
(B-L)_{\rm extra} = - (N_Q + N_U + N_D) + 3 (N_L + N_N + N_E),
\end{equation}
where `extra' refers to fourth generation particles.  The decay of the
right-handed neutrinos generates only $L$, while both $B$ and
$(B-L)_{\rm extra}$ remain vanishing. $Y$ also remains vanishing by
gauge invariance. Since the number densities of the various species are
proportional to their chemical potentials at the lowest order, we can
derive nontrivial relations by considering the constraints imposed by
chemical equilibrium.

The chemical equilibrium due to the Yukawa interactions implies
\begin{eqnarray}
& & \mu_q = \mu_u + \mu_H = \mu_d - \mu_H,\\
& & \mu_l = \mu_e - \mu_H,\\
& & \mu_Q = \mu_U + \mu_H = \mu_D - \mu_H,\\
& & \mu_L = \mu_N + \mu_H = \mu_E - \mu_H,
\end{eqnarray}
while the electroweak anomaly effect requires
\begin{equation}
9\mu_q + 3\mu_l + 3\mu_Q + \mu_L = 0.
\end{equation}
Here, $\mu_i$ refers to the chemical potential of a particle of
species $i$, and $H$ is the standard electroweak Higgs boson. Then we find
\begin{eqnarray}
B &\propto& \frac{1}{3} (18\mu_q + 9 \mu_u + 9 \mu_d)
      - (6\mu_Q + 3\mu_U + 3\mu_D) = 0,\\
(B-L)_{\rm extra} &\propto& - (6\mu_Q + 3\mu_U + 3\mu_D)
      + 3 (2\mu_L + \mu_N + \mu_E) = 0,\\
Y &\propto& \frac{1}{6} (18\mu_q + 6\mu_Q)
      + \frac{2}{3} (9\mu_u + 3\mu_U)
      - \frac{1}{3} (9\mu_d + 3\mu_D) \nonumber \\
& &   - \frac{1}{2} (6\mu_l + 2\mu_L)
      - (3\mu_e + \mu_E) - 2\mu_H = 0,\\
L &\propto& (6\mu_l + 3\mu_e) - 3 (2\mu_L + \mu_E + \mu_N) \neq 0.
\end{eqnarray}
Solving these constraints, we obtain
\begin{eqnarray}
N_q + N_u + N_d &=& - \frac{108}{137} L,\label{eq:a1}\\
N_l + N_e &=& \frac{101}{137} L,\label{eq:a2}\\
N_Q + N_U + N_D &=& - \frac{36}{137} L,\label{eq:a3}\\
N_L + N_E + N_N &=& - \frac{12}{137} L.\label{eq:a4}
\end{eqnarray}
Thus, we see explicitly that the non-vanishing lepton number generated
by the decay of the right-handed neutrinos will be partially converted
to a nonvanishing baryon number for particles from the ordinary families,
as well as nonvanishing baryon and lepton numbers for particles
from the extra family.

One potential cosmological problem with this scenario is that the
particles from the extra family could overclose the Universe. The
constraints from primordial nucleosynthesis imply that baryons in the
ordinary families must have a present cosmic density in the range
$\Omega_b h_0^2 = 0.010$--$0.15$, where $h_0=0.4$--$1$ is the reduced
Hubble constant \cite{Walker}. On the other hand, the quarks and leptons
in the extra family have also acquired an asymmetry that will
remain until the present. Based on the predicted ratio of these
asymmetries, the new contributions to the cosmic density are
\begin{eqnarray}
\Omega_{Q,U,D} &=& \frac{m_F}{m_p} \Omega_b,\\
\Omega_{L,E,N} &=& \frac{1}{3} \frac{m_F}{m_p} \Omega_b,
\end{eqnarray}
where $m_p$ is the mass of the proton. Even in the extreme case where
$\Omega_b = 0.01$, the fourth generation particles would overclose the
Universe when $m_F \gtrsim 100$~GeV.   One might hope that these
fourth generation particles could be candidates for cold dark matter.
However, there are very strong observational constraints against
dark matter that is strongly interacting \cite{SIMP}, charged
\cite{CHAMP} or composed of Dirac neutrinos \cite{Caldwell}.

Fortunately, this problem can be avoided because baryon number is
spontaneously broken, and we can find a way to make the fourth
generation particles decay.  Suppose that U(1)$_B$ is broken by an
electroweak-singlet Higgs field $\chi$ with the following quantum
numbers under SU(3)$_C \times$ SU(2)$_L \times$ U(1)$_Y \times$
U(1)$_B$:\footnote{This charge assignment can be embedded into SU(4)$_H$
{\bf 15} representation, which allows the operator in Eq.~(\ref{eq:dfo}).}
\begin{displaymath}
\chi ({\bf 1, 1, 1})_{\bf +4/3} .
\end{displaymath}
Then the following dimension-five operators are the only ones for the
quarks that are consistent with the gauge symmetry below $M_H$,
\begin{equation}
{\cal L}_5 = \frac{1}{M_V} (\bar{q} \, U H \chi
      + \bar{Q} u H \chi^*
      + \bar{q} D H^* \chi
      + \bar{Q} d H^* \chi^*) + h.c.
\label{eq:dfo}
\end{equation}
(Of course, there are similar operators involving the lepton fields.)
One could imagine that these operators are generated by the exchange
of a heavy vector-like quark with mass $M_V$.  As a consequence of
Eq.~(\ref{eq:dfo}), particles in the extra family can decay into
ordinary particles.
The decay rate is given by
\begin{equation}
\Gamma_F \sim \frac{1}{8\pi}
      \left( \frac{v\langle\chi\rangle}{M_V m_F} \right)^2 m_F ,
\end{equation}
where $v=246$~GeV is the expectation value of the electroweak Higgs $H$,
and $\langle \chi \rangle$ is the scale of baryon number symmetry breaking
which we assume is around the electroweak scale. It is now clear that the
particles in the extra family can decay well before nucleosynthesis as long
as $M_V \lesssim 10^{15}$~GeV. Note that the decay of particles in the
extra family gives an additional contribution to the cosmic baryon
asymmetry of the ordinary particles that does not cancel out the original
asymmetry that we obtained in Eq.~(\ref{eq:a1}).

The particles (especially quarks) in the fourth generation could be
produced at the Tevatron or LHC. Their signatures depend on whether they
leave the detector before or after they decay. If $M_V \lesssim
10^{11}$~GeV, they decay inside the detector and leave a signature
similar to that of the top quark.  In this case, the fourth generation
fermions must have masses larger than $\sim 140$~GeV
\cite{D0top}. If $M_V$ is larger, they could be detected before they
decay. A search for stable color-triplet quarks was carried out by CDF
\cite{CDF}, but the present constraint is rather weak (50--116~GeV ruled
out at 95~\% C.L.).

An important point in this model is that proton decay is forbidden
even though baryon number is spontaneously broken.
Any baryon-number violating effects can be described in terms of
effective operators with powers of the order parameter $\langle
\chi \rangle$ which breaks U(1)$_B$. The general structure of
such an operator is ${\cal O} = q^k l^m \chi^{*n}$, where $q$ is a quark
field, $l$ lepton, and $k,m,n$ are integers. Since lepton fields carry
integer charges and $\chi$ neutral, the power $k$ has to be a multiple
of three, $k = 3l$ where $l$ is another integer. Then the factor $q^k$
carries baryon number $l$ which has to be compensated by the baryon
number of $\chi$, $-(4/3)n$. Therefore $n$ also has to be a multiple of
three, $n = 3p$, and the operator has the form
\begin{equation}
{\cal O} = (q^{12} \chi^{*3})^p l^m  ,
\end{equation}
which has dimension $21 p + (3/2) m$.\footnote{This operator can be
written in an explicitly SU(2)$_L \times $U(1)$_Y$ symmetric way: {\it
e.g.}\/ for $m=0$, $p=1$, ${\cal O} = (q \cdot q)^4 d^4 \chi^{*3}$.}
Not only is this operator extremely suppressed by powers of a high
mass scale, it also cannot contribute to proton decay because the
quark field is raised to a power that is too large.  Thus, there
is no perturbative contribution to proton decay in this
model. There could be a contribution from the electroweak instanton
effect, but the decay rate due to the anomaly is known
to be extremely tiny \cite{tHooft}. Therefore, nucleon decay is
effectively forbidden in this model.

Finally, one might worry that $\gamma_B$ exchange may lead to
flavor-changing neutral current because it is coupled to a matrix
$\mbox{diag}(1, 1, 1, -3)$ in the flavor space of the model.
Mixing between the ordinary and extra families gives rise to off-diagonal
coupling for the $\gamma_B$. However as seen above, the mixing is suppressed
by a power of $\langle \chi \rangle/M_V$, and the off-diagonal coupling
between different generations by a square of this suppression factor.
All constraints from flavor-changing neutral currents are avoided when
$M_V \gtrsim 100$~TeV. A similar lower bound applies to the mass of the
horizontal gauge bosons in SU(3)$_H$.

\section{Leptonic Signals}
\setcounter{equation}{0}

We have shown that there is a class of models with gauged baryon number
in which the kinetic mixing between the hypercharge and baryon number
gauge bosons is naturally small below the electroweak scale.
Nevertheless, a small amount of mixing is not necessarily a bad thing,
because it can provide us with a possible leptonic signature for our model.
In this section we consider the new contribution to the Drell-Yan
production of lepton pairs at hadron colliders.  In particular, we show
that the signal may be within the reach of an upgraded Tevatron.

The quantity of interest is $d\sigma/dM$, the differential cross section as
a function of the lepton pair invariant mass.  One can obtain the desired
result from the conventional expression for the Drell-Yan differential cross
section by making the substitutions
\beq
g_V^i \rightarrow g_V^i + \frac{2}{3} c_Z c_w s_w^2 \sqrt{\frac{\alpha_B}
{\alpha}}\frac{\hat{s}}{\hat{s}-m_B^2+i m_B \Gamma_B}
\eeq
and
\beq
Q^i \rightarrow Q^i - \frac{1}{3} c_\gamma c_w \sqrt{\frac{\alpha_B}
{\alpha}} \frac{\hat{s}}{\hat{s}-m_B^2+i m_B \Gamma_B} ,
\eeq
where $\hat{s}$ is the parton center of mass energy squared.  Here $Q^i$
is the the quark charge in units of $e$, and $e g_V^i/ (2 c_w s_w)$ is
the vector coupling of the $Z$ to a quark of flavor $i$, with $g_V^i =
T_{3L}-2Q^i s_w^2$.

Our results for $d\sigma/dM$ in a $p\overline{p}$ collision at
$\sqrt{s}=1.8$ TeV are shown in Fig.~4 for one lepton species,
integrated over the rapidity interval $-1<y<1$, using the EHLQ Set II
structure functions \cite{ehlq}. This range in rapidity was chosen to be
consistent with the CDF detector coverage \cite{cdfdy}.  The solid line
shows the conventional differential cross section, (with
$c_\gamma=c_Z=0$), while the dotted lines give our results for
$c_\gamma=c_Z=0.01$.  For the values of $m_B$ shown, the results do not
depend strongly on the precise choice for $c_Z$.  Around the $\gamma_B$
mass there is a noticeable excess of events beyond the expected
background.  Because this excess is an interference effect, it depends
linearly on $c_\gamma$.  We show the excess in the total dielectron plus
dimoun signal in a bin of size $dM$ surrounding the $\gamma_B$ mass in
Table~2, for $m_B=30$, $40$, and $50$ GeV. The statistical significance
of the signal assuming integrated luminosities of 1 fb$^{-1}$ and 10
fb$^{-1}$ is also shown.  The largest excess at 1 fb$^{-1}$, is a 5.4
standard deviation effect for $m_B=30$ GeV. However, with 10 fb$^{-1}$
of integrated luminosity, even the excess at 50 GeV would be detectable
at the 9.4 sigma level. This simple analysis is sufficient for a
qualitative understanding of the signal we might expect to find at the
Tevatron, with both the main injector, and a luminosity upgrade. We have
not included the efficiency of the cuts and acceptance, but it is rather
high even in a realistic analysis (93~\% for $e^+ e^-$ and 82~\% for
$\mu^+ \mu^-$ in CDF analysis \cite{cdfdy}).  A more exhaustive
study, including the efficiency of the cuts and detector acceptance,
as well as a comparison of the shape of the differential cross section to
that expected in our model is required for a more accurate assessment of the
discovery potential for this model at the Tevatron.  It is interesting
to note that even if the coupling $\alpha_B$ is smaller than 0.1, and the
jet physics discussed in Ref.~\cite{carmur1} is no longer of relevance, the
mixing effect that we discuss here could still be significant enough to
provide a clear signal for the model.

\begin{table}
\caption[]{Excess Dielectron plus Dimuon Production at the Tevatron.}
\begin{center}
\begin{tabular}{cccc|cc}
\hline
$m_B$ & $dM$ & Background & Excess
	& \multicolumn{2}{c}{statistical significance}\\ \cline{5-6}
(GeV) &(GeV) &   (fb)    & (fb)   & 1~fb$^{-1}$ & 10~fb$^{-1}$\\
\hline\hline
30 & 2 & 3468 & 320 & 5.4~$\sigma$ & 17.2~$\sigma$ \\
40 & 4 & 2798 & 208  & 3.9~$\sigma$ & 12.4~$\sigma$ \\
50 & 4 & 1422  & 112  & 3.0~$\sigma$ & 9.4~$\sigma$ \\
\hline\hline
\end{tabular}
\end{center}
\end{table}

\section{Conclusions}
\setcounter{equation}{0}
We have shown that there are models with gauged baryon
number in which kinetic mixing between the baryon number and
hypercharge gauge bosons is naturally absent above the electroweak scale.
Since the mixing is generated only through radiative corrections
at lower energies, the resulting effective theory is consistent
with precision electroweak measurements even when $\alpha_B$ is as
large as $0.1$, as we showed quantitatively
in section~2.  The exciting feature of the type of models that we proposed
is that the baryon number gauge boson $\gamma_B$ can be lighter
than $m_Z$ with a large gauge coupling, and yet be hidden in existing
LEP and Tevatron data.  This is the point that we emphasized
in Ref.~\cite{carmur1}. However, even if the the coupling $\alpha_B$ is
not large enough to produce an unambiguous hadronic signal, we have shown
that the kinetic mixing term may give us another means for detecting
the $\gamma_B$ via its contribution to Drell-Yan dilepton production
at hadron colliders.  With both the main injector and a luminosity upgrade,
this signal may eventually be within the reach of the Fermilab Tevatron.

\begin{center}
{\bf Acknowledgments}
\end{center}
We are grateful to Lawrence Hall for useful comments. We thank
Don Groom from the Particle Data Group for providing us with
compiled data on $R$.
{\em This work was supported by the Director, Office of Energy Research,
Office of High Energy and Nuclear Physics, Division of High Energy
Physics of the U.S. Department of Energy under Contract DE-AC03-76SF00098.}



\newpage
\begin{center}
{\bf Figure Captions}
\end{center}

{\bf Fig.~1.}  Running of $c_\gamma$ and $c_Z$, assuming
$c_\gamma (\Lambda) =c_Z (\Lambda) =0$ at $\Lambda= 250$ GeV.

{\bf Fig.~2.}  Constraints on $c_Z(m_Z)$ from the two standard
deviations of the
experimental uncertainties in the $Z$ mass, hadronic width, and
$Z\rightarrow b\overline{b}$ forward-backward asymmetry.

{\bf Fig.~3.}  Constraints on $c_\gamma(m_B)$ from the two standard
deviations of the experimental uncertainty in $R$ measured at various
$\sqrt{s}$
as compiled by the Particle Data Group.   The running of $c_\gamma$
corresponding to $\Lambda=200$ GeV is shown for comparison.

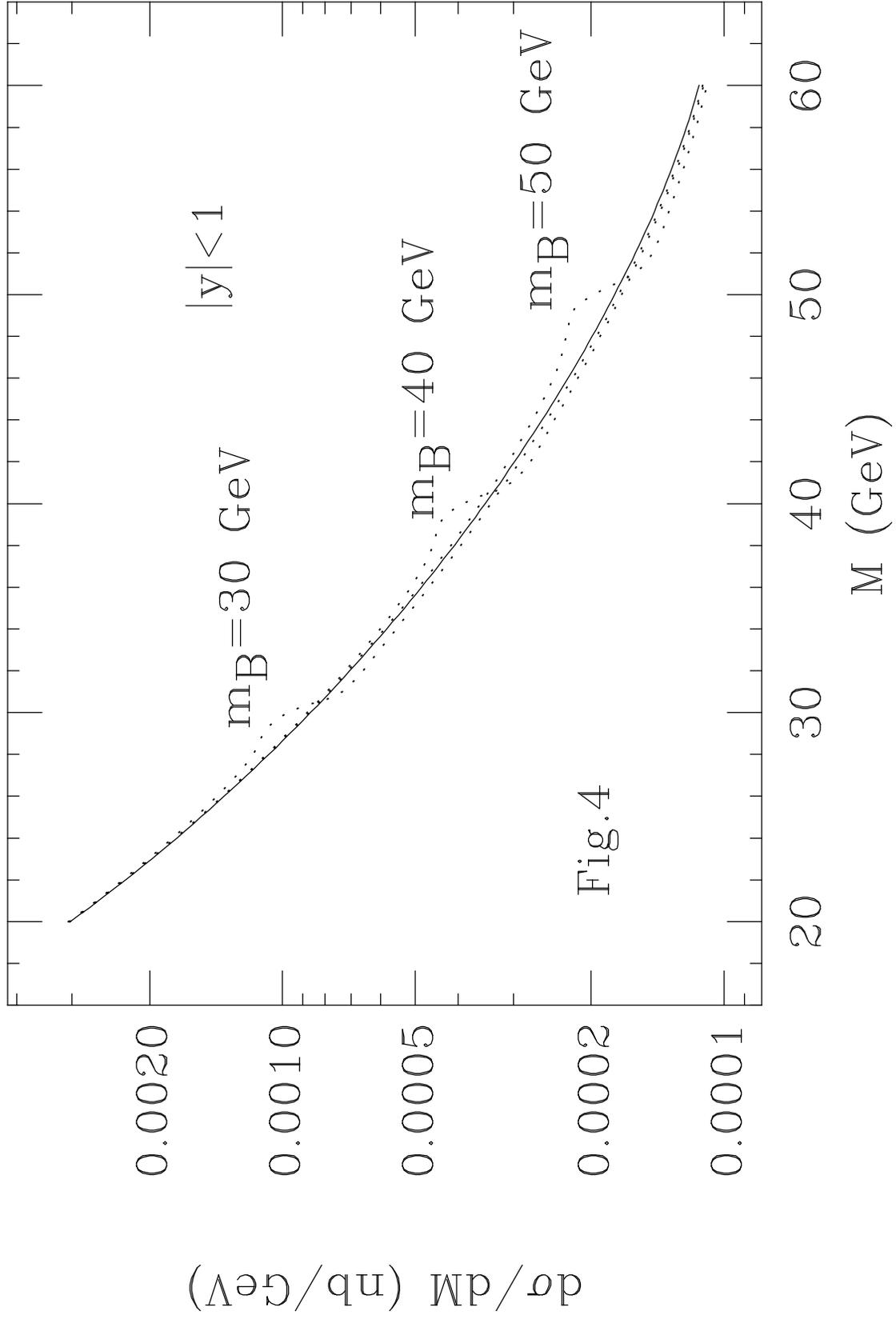
{\bf Fig.~4.} Drell-Yan dilepton differential cross section as a
function of the lepton pair invariant mass, integrated over the
rapidity interval $|y|<1$, for one lepton species. The dashed curves
include the effect of $\gamma_B$ exchange, assuming
$c_\gamma (m_B) = c_Z (m_B) = 0.01$, for $m_B=30$, $40$, and
$50$ GeV, respectively.

\end{document}